\documentclass[letter]{aa}
\usepackage{txfonts}
\usepackage{natbib}
\usepackage{graphicx}
\bibpunct{(}{)}{;}{a}{}{,}
\begin{document}

\title{Is a binary fraction-age relation responsible for the
lack of EHB binaries in globular clusters?
\thanks{Based on observations with the ESO Very Large Telescope
at Paranal Observatory, Chile (proposal ID 075.D-0492(A))}
}

\author{
C. Moni Bidin \inst{1}
\and
M. Catelan \inst{2}
\and
M. Altmann \inst{1,3}
}

\institute{
Departamento de Astronom\'{i}a, Universidad de Chile,
Casilla 36-D, Santiago, Chile
\and
Pontificia Universidad Cat\'{o}lica de Chile, Departamento de Astronom\'{i}a
y Astrof\'{i}sica, Av. Vicu\~{n}a Mackenna 4860, 782-0436 Macul, Santiago, Chile
\and
University of Heidelberg, Centre for Astronomy, M\"{o}nchhofstr. 12-14,
D-69120 Heidelberg, Germany
}
\date{Received / Accepted }


\abstract{The recently-discovered lack of close binaries, among extreme horizontal
branch (EHB) stars in Galactic globular clusters, has thus far constituted a major
puzzle, in view of the fact that blue subdwarf stars~-- the field counterparts
of cluster EHB stars~-- are well-known to present a high binary fraction.}
{In this {\em Letter}, we provide new results that confirm the lack of close
EHB binaries in globular clusters, and present a first scenario to explain the
difference between field and cluster EHB stars.}
{First, in order to confirm that the lack of EHB binaries in globular clusters
is a statistically robust result, we undertook a new analysis of 145 horizontal
branch stars in NGC\,6752, out of which forty-one belong to the EHB. To search for
radial-velocity variations as a function of time, we repeated high-resolution
($R=18\,500$) spectroscopy of all stars, four times during a single night of
observations.}
{We detected a single, hot (25\,000~K), radial-velocity variable star as a
close-binary candidate. From these results, we estimate
an upper-limit for the close (period $P \leq 5$~day) binary fraction $f$ among
NGC\,6752 EHB stars of 16\% (95\% confidence level), with the most probable
value being $f=4\%$. Thus our results clearly confirm the lack of
close binaries among the hot HB stars in this cluster.}
{We suggest that the confirmed discrepancy between the binary fractions for field
and cluster EHB stars is the consequence of an
$f$-age relation, with close binaries being more likely in the case of younger
systems.
We analyze theoretical and observational results available
in the literature, which support this scenario.
If so, an age difference between the EHB progenitors in the field and
in clusters, the former being younger (on average) by up to several Gyr, would
naturally account for the startling differences in binary fraction between the
two populations.
}

\authorrunning{}
\maketitle


\section{Introduction}
\label{capintro}

The presence of a large population of binaries among field B-type subdwarf (sdB) stars,
also referred to as Extreme Horizontal Branch (EHB) stars,
is well-established in the literature \citep{Ferguson84,Allard94,Ulla98,Aznar01,Maxted01,
Williams01,Reed04,Napiwotzki04}. The measurement of binary fraction varies from one survey
to another, probably because of the different survey selection effects.
It is generally agreed however, that the binary fraction must be large. Moreover, extensive
analysis of orbital parameters \citep{Moran99,Saffer98,Heber02,MoralesRueda04,MoralesRueda06}
reveal that {\em close} binaries ($P \leq 10$~d) comprise
a major fraction of the field EHB stars, with short-period systems---including either a
degenerate or a low-mass main sequence (MS) companion---constituting about half of the
entire field sdB population. Indeed, theoretical results indicate that these stars can
be naturally explained within the context of binary star evolution \citep{Han02,Han03}.
Moreover, EHB stars of binary origin may also account for the UV flux excess (``UV upturn'')
observed in elliptical galaxies \citep{Han07}, although it should be noted that
single-star scenarios have also been suggested \citep[see][for a recent review]{Catelan07}.
One way or another, the link between field sdB stars and binary systems is very
well-established, both on observational and theoretical grounds, and close, short-period
systems are the most frequently found amongst them.

It therefore came as a great surprise that the first radial-velocity (RV) surveys among
EHB stars in globular clusters (GCs) revealed a remarkable lack of close binary systems
\citep[see][for a review]{Moni07b}.

In this {\em Letter} we present, in light of new observational results, a first scenario
to account for this discrepancy.


\begin{figure*}[t]
\begin{center}
\includegraphics[width=17cm]{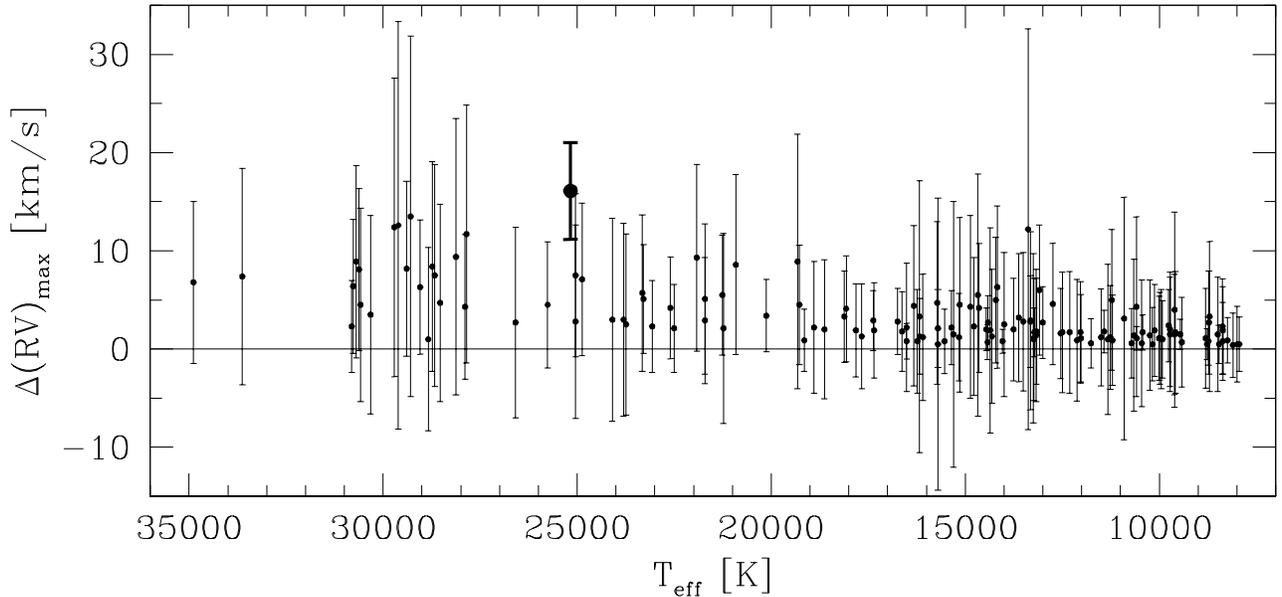}
\caption{Results of our RV variability search. The maximum RV variation
(${\rm RV}_{\rm max}-{\rm RV}_{\rm min}$)
observed for each star is plotted at its effective temperature. For the sake of
clarity, we plot only the 3$\sigma$ error bars, which indicate the statistical significance
of the observed variations. The data for the only star with a clearly-detected RV variation,
is plotted using a larger, solid circle symbol, and thicker lines to indicate its errorbar.}
\label{figresults}
\end{center}
\end{figure*}

\section{Observations and measurements}
\label{capdata}

High-resolution spectra ($R=18\,500$) centered on H$_\mathrm{\beta}$
(4700--4970 \AA) were collected with 3600s exposures on June 29,
2005 at VLT-UT2, with the FLAMES-GIRAFFE optical spectrograph in MEDUSA
configuration (setup HR7A).
169 HB stars were selected from the \citet{Momany02} 
photometry, spanning the whole temperature range along the HB, out of which 54 were
EHB stars (i.e., with $T_\mathrm{eff}\geq 20\,000$~K).
When performing the target selection, we ensured that each individual candidate
was sufficiently isolated that no other bright star fell within the fiber,
extending observations as far as possible
toward the cluster center. Selected targets were divided into two fiber 
configurations. In Table \ref{tabtime}, we indicate the time at the start of
each exposure, in hours from the first spectrum.
To confirm that our targets were true cluster HB stars, we verified
that absolute radial velocities are consistent with cluster membership, and we
inspected all spectra by eye. We excluded six, cool-star targets
(T$_\mathrm{eff}\approx 6\,500$~K) from our analysis,
because their cluster membership was doubtful.

All data reduction steps were completed using ESO's GIRAFFE pipeline.
Wavelength calibration was performed using lamp fiber data acquired simultaneously
during the observations: according to the FLAMES manual
the calibration error is about 0.15~km~s$^{-1}$.

\begin{table}[t]
\begin{center}
\caption{Time elapsed (in hours) between the start of each exposure and the 
first one, for each fiber configuration (Medusa~1 and Medusa~2).}
\label{tabtime}
\begin{tabular}{c| c c}
\hline
\hline
frame & Medusa1 & Medusa2 \\
 & \multicolumn{2}{|c}{hours} \\
\hline
1 & 0.00 & 0.00 \\
2 & 1.03 & 4.73 \\
3 & 4.75 & 5.77 \\
4 & 5.78 & 9.13 \\
\hline
\end{tabular}
\end{center}
\end{table}

RV's were measured using the cross-correlation (CC) technique \citep{Tonry79},
cross-correlating each spectrum with synthetic templates of appropriate temperature 
and gravity from the library of \citet{Munari05}.
During CC we used only H$_{\beta}$ with full wings, neglecting all weaker lines
that were hardly visible due to noise, and highly variable from star to star.
Some tests carried out prior to the actual measurements revealed
that, with this limitation, the adopted metallicity of the template had a negligible
impact on the results.
Errors in the measured RV variations were calculated as the quadratic
sum of the errors for the individual measurements.

More details about the observations, data reduction and measurements will be
provided in a forthcoming paper (Moni Bidin et al., {\em in preparation}).


\section{Results}
\label{capresults}

We were unable to acquire observations for twelve stars, due to for example, light
contamination by nearby fibers, too low signal-to-noise ratio, or the star being
badly-centered in the fiber. Moreover, for six
of the hotter stars one or more spectra were of too low S/N to provide reliable measurements;
the observations being thus incomplete for these stars, we excluded them from subsequent 
analysis. Therefore, we successfully performed a RV analysis for 145 stars, out of which
41 fall along the EHB.

Our analysis revealed that the errors given by CC alone were underestimated. We evaluated 
this effect by analyzing statistically the distribution of $\frac{\Delta(\mathrm{RV})}{\sigma}$ 
for the cooler stars ($T_\mathrm{eff}\leq 10\,000$~K). Among this subsample, we do not expect
a population of very close binaries, and
observed variations can be ascribed only to random errors. We found that
the latter were underestimated by a factor of 1.44.
We expected this factor to increase for decreasing S/N, but repeating the analysis
in different temperature bins we found only small variations. Moreover,
evaluating the correction factor for hotter stars would invalidate our results,
because RV variations due to binarity, if not properly identified,
would spuriously enlarge the estimated errors and thus incorrectly lead to the lack
of detected binaries.
Hence we applied the correction by a factor 1.44 to the whole sample, as a good approximation
valid for all temperatures.

Our results are summarized in Figure \ref{figresults}. For each star, we plot the
maximum RV variation and the 3$\sigma$ interval. Effective temperatures were estimated
using the color-T$_\mathrm{eff}$ relation derived in \citet{Moni06a}.
Because each datum is the maximum of six measurements, the probability that each datum exceeds
3$\sigma$ is 1.6\%, and we expect 0.66 false detections out of 41 EHB targets.
We adopted this threshold to select objects displaying
suspect variations to be further analyzed, but we found no need to
discuss border cases because the only candidate found
shows variation at the 10$\sigma$ level (16.1$\pm$1.6 km s$^{-1}$).
This star was already highlighted by \citet{Moni07} for its red color and evident
Mg\,Ib triplet, signatures of a cool MS companion. We are currently analyzing data for
this star, which turns out to be the first sdB+MS close system ever found in a globular cluster.
In summary, we come out with only one good close binary candidate in our whole sample of
145 HB (41 EHB) stars.

\begin{figure}[t]
\begin{center}
\resizebox{\hsize}{!}{\includegraphics{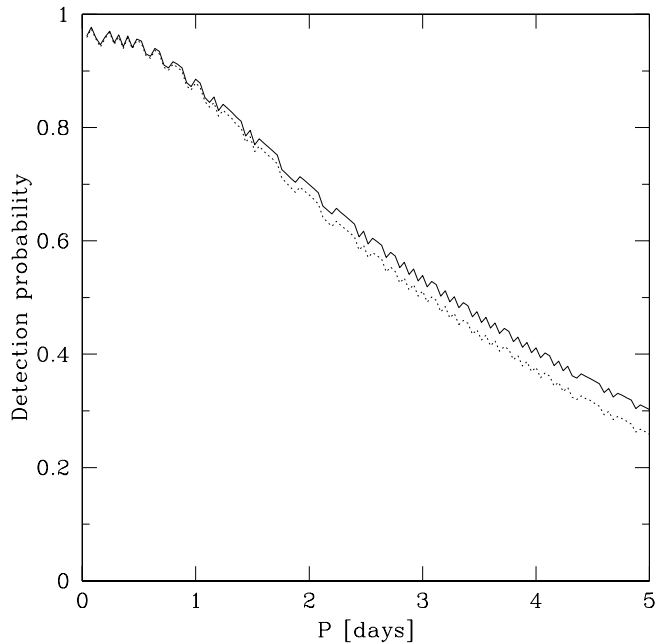}}
\caption{Probability of detecting a close binary systems in our survey, as a function of its period.
As probability varies with both times of observations and errors in the measured RV variations, we 
plot both the average over the probabilities for each individual star ({\em solid line}) and the 
one calculated assuming a fixed error of $10\,{\rm km\,s}^{-1}$ for all stars ({\em dotted line}).}
\label{figprob}
\end{center}
\end{figure}

In Figure~\ref{figprob}, we plot the estimated detection probability in our survey as
a function of the period. This was calculated as in \citet{Moni06a}, assuming a mass 
of $0.5\, M_{\sun}$ for both components and circular orbits, and
integrating over the phase and the orbital inclination angle.
We derived the detection probability for each
individual star (given sampling times in Table~\ref{tabtime} and the aforementioned
3$\sigma$ values), and then averaged it to obtain the final curve. We also checked 
that other reasonable ways of tackling the problem (for example, assuming a fixed 
$3\sigma = 10\,{\rm km\,s}^{-1}$ for all stars in the sample) did not change our  
results by more than 1-3\%. Clearly, the detection probability rapidly falls down
for increasing periods, as a consequence of the observations being restricted to a
single night; accordingly, we will limit our analysis to systems with $P \leq 5$~d.
It must be emphasized that such a limitation is not particularly severe, since many
surveys have revealed that the bulk of the field binary sdB population has periods
of approximately
$P \approx 1$~d, systems with $P \geq 5$~d constituting only the tail end of the period
distribution \citep[see, for example,][]{MoralesRueda03}.

Using the calculated detection probability, we estimated the close binary fraction
$f_{P\leq 5\,{\rm d}}$ among EHB stars as in \citet{Maxted01}, assuming a flat distribution
of periods, and the successful detection of one binary out of 41 targets. We calculated
$f_{P\leq 5\,{\rm d}}$ assuming a Gaussian distribution in $\log P$, as in
\citet{Maxted01} and \citet{Napiwotzki04}, which better represents the period distribution 
for field sdBs---but the results do not differ noticeably.
The derived probability distribution peaks at $f = 4\%$. It is markedly non-Gaussian,
but it falls below 5\% for $f\geq 16\%$.
We conclude that the best estimate (i.e., the most likely value)
for $f_{P\leq 5\,{\rm d}}$ is 4\%, and that $f_{P\leq 5\,{\rm d}}\leq$16\% at
the 95\% confidence level. Our derived upper limit is in perfect agreement with the one
that was found for $f_{P\leq 10\,{\rm d}}$ by \citet{Moni06a}. We note that these results are
at variance with the preliminary ones previously obtained by \citet{Peterson02}, who
claimed the detection of many close binaries among the EHB stars in the same cluster.
The reader is referred to \citet{Moni06a} for a discussion of this discrepancy.


\section{Is there a close binary fraction-age relation?}
\label{capdiscussion}

Using an independent dataset, a sample more than twice as large as in \citet{Moni06a},
and a resolution in RV variations that is higher by almost a factor of two, for the first
time we were able to find a good binary candidate among the EHB stars in
\object{NGC\,6752}.
That notwithstanding, our results confirm that the corresponding (close) binary fraction $f$ is
very small in \object{NGC\,6752}, with a most likely value of
$f = 4\%$ and an upper limit of $f = 16\%$ at the 95\% confidence level.

There are hints that a small $f$ is not a peculiarity of \object{NGC\,6752}, but
could also be a characteristic of other globular clusters \citep{Moni06b}.
This is in sharp contrast with
the situation for field sdB stars, where close binaries are at least a factor of ten more
frequent, comprising up
to 70\% of the entire sdB population \citep[see \S4 in][for a recent review]{Catelan07}.
There is however no knowledge about cluster EHB stars in long-period binaries, or
with a close low-mass companion, because no survey has investigated their role yet.
These kind of systems are known to exist among field sdBs, but are just a minor population,
and their presence in GCs would not alleviate the striking
contrast with field results.

What is the origin of this startling difference between field and cluster EHB stars?
We believe that it may not be completely unexpected: there should be
a relation between the close binary fraction and the mean age of a sdB population,
as a consequence of the different efficiency of binary channels responsible
for EHB star formation.

Theoretical arguments strongly suggest
that sdB stars in close binary systems should have undergone at least one common envelope
(CE) phase. Although \citet{Han02} found an upper limit for the initial mass of the sdB
progenitor in this scenario, they also pointed out that within the permitted values a
higher initial mass favors the CE channel, and leads to sdB binaries with shorter periods.
In fact, a higher mass implies a more tightly-bound envelope, which requires a greater
amount of (orbital) energy to be released. On the other hand, \citet{Han02} explored 
the stable Roche Lobe Overflow (RLOF) scenario, and found that a higher initial
progenitor mass makes it harder for the RLOF to be stable (see their Table~3), because 
the minimum mass of the companion increases with increasing progenitor mass. For higher 
values, fewer MS and white dwarf (WD) secondaries are sufficiently massive for the
activation of this channel (sdB's with neutron star companions are indeed very rare).
The progeny of systems that underwent stable RLOF, that is wide binaries with very long 
periods, is therefore generated primarily by progenitors of lower initial mass.

In light of these results, a relation between $f$ and the mean initial mass of the sdB 
progenitors could be naturally expected, hence implying a relationship between
$f$ and (mean) age.
More specifically, one may naturally expect that field sdB's formed from progenitors with
a wide spectrum of initial masses (up to about $2\,M_{\sun}$), whereas in an old population 
(such as in globular clusters) only the progeny of less massive stars are currently found
on the EHB, those of more massive ones having long evolved away from the He-burning phase.
The stable RLOF is an efficient channel for sdB formation in the old case, while the 
CE one is not---and the CE itself would be released at earlier stages, before the 
orbits shrink substantially. Therefore, in old populations we should expect to find 
predominantly wide binaries, or/and single EHB stars formed through other channels 
\citep[see][for a recent discussion]{Catelan07}.
In fact, WD mergers, the third binary channel studied by \citet{Han02},
can form (single) sdB stars, and is expected to be particularly efficient
for old populations (see below). Moreover, it may be expected to
play an important role within the dense environment of a globular cluster,
where stellar encounters can harden close binaries \citep{Heggie75} and thus enhance
channel efficiency, as proposed earlier by \citet{Bailyn89}.
However, dynamical effects are not required
for the formation of EHB stars \citep{Whitney98}, as indicated by the lack of
radial gradients among the EHB stars in \object{NGC\,2808} \citep{Bedin00}
and $\omega$~Centauri \citep[\object{NGC\,5139};][]{DCruz00}.

There are other results in the recent literature that provide support to our proposed
scenario. In fact, \citet{Napiwotzki04} already invoked a possible $f$-age or $f$-metallicity 
relation to explain their results. Among field sdB stars, they found a close binary fraction 
lower than in previous surveys, and noted that their sample was on average fainter, possibly
including more thick disk or halo (hence older/metal poorer) stars.
This suggested difference between the data used has never been fully investigated,
and a further check of the populations of these stars would be invaluable.
Even stronger support is provided by the recent simulations
by \citet{Han07}, modeling the UV upturn of elliptical galaxies from a binary sdB population.
Their Figure~7, though aimed at reproducing the spectral energy distribution of elliptical 
galaxies, shows how the relative contribution of sdB's formed through different channels 
evolves with time. One finds that the contribution to the UV flux from RLOF sdB's is
essentially constant with time, whereas the one from sdB's that experienced the CE phase 
first becomes important for a population of age 1.5~Gyr and then slowly fades,
being more and more marginal for increasing ages. On the other hand,
for populations older than 5~Gyr these authors find that sdB's formed through the merger 
channel dominate---but these are not binaries any longer.
We suggest that these theoretical results implicitly hint at a solution to the 
heretofore puzzling lack of close EHB binaries in globular clusters, on the one
hand, and their large numbers among field stars, on the other.

As an alternative to the binary scenario,
it is worth noting that many single-star evolutionary channels
have been invoked to explain EHB star formation in GCs, including
interactions with a close planet \citep[][see also \citealt{Silvotti07}]{Soker98},
He mixing driven by internal rotation \citep{Sweigart79,Sweigart97} or by stellar
encounters \citep{Suda07}, dredge-up induced by
H-shell instabilities (\citealt{vonRudloff88}, but see also \citealt{Denissenkov03}),
close encounters with a central, intermediate-mass black hole \citep{Miocchi07}, and
a sub-population of stars with high helium abundance \citep[e.g.,][]{DAntona05}.

We note that any EHB formation mechanism involving single progenitors should be
particularly inefficient among field stars (but possibly not so among GCs), since 
the EHB progenitors are on average younger, and thus more massive, in the field than 
in GCs. As a consequence, a much higher amount of RGB mass loss is needed to successfully 
produce a EHB star in the case of a field progenitor, whereas the ancient GC red giants 
have a much smaller envelope to be removed before they become EHB stars. This 
notwithstanding, a small component of single-star progeny may still be needed 
among field stars to fully explain the available observations \citep{Lisker05}.


\begin{acknowledgements}

We warmly thank Yazan Momany for providing us with his accurate astrometric data.
MC is grateful to Allen V. Sweigart for insightful discussions on the origin
of hot HB stars. CMB acknowledges Universidad de Chile graduate fellowship
support from programs MECE Educaci\'{o}n Superior UCH0118 and Fundaci\'{o}n Andes
C-13798. Support for MC is provided by Proyecto Fondecyt Regular \#1071002.
MA was funded by FONDAP 1501 0003.

\end{acknowledgements}


\bibliographystyle{aa}
\bibliography{biblioGirI}

\end{document}